\documentclass[conference]{IEEEtran}

\usepackage{graphicx}
\usepackage{latexsym}
\usepackage{amsfonts}
\usepackage{amssymb}
\usepackage{amsbsy}
\usepackage{amsmath}
\usepackage{multicol}
\usepackage{float}
\usepackage{subfigure}
\usepackage{algorithm}
\usepackage[dvips]{epsfig}

\usepackage{fancyhdr}
\usepackage{epsfig}
\usepackage{threeparttable}
\usepackage{epsf,epsfig}
\usepackage{amsthm}
\usepackage[noadjust]{cite}
\usepackage{dsfont}
\usepackage{color}
\usepackage{enumerate}

\newtheorem{lemma}{Lemma}
\newtheorem{proposition}{Proposition}
\newtheorem{corollary}{Corollary}

\def\E{\mathsf{E}}

\def\SIR{\mathsf{SIR}}

\def\({\left(}
\def\){\right)}
\def\[{\left[}
\def\]{\right]}

\def\nn{\nonumber}

\usepackage{dblfloatfix} 

\def\papertitle{User Attraction via Wireless Charging\\ in Downlink Cellular Networks}

\begin{document}

\title{ \fontsize{28}{10}\selectfont \papertitle}

\author{\IEEEauthorblockN{ Jeemin Kim, Jihong Park, Seung-Woo Ko, and Seong-Lyun Kim}
\IEEEauthorblockA{School of Electrical and Electronic Engineering, Yonsei University\\
50 Yonsei-Ro, Seodaemun-Gu, Seoul 120-749, Korea\\
Email: \{jmkim, jhpark.james, swko, slkim\}@ramo.yonsei.ac.kr}}


\maketitle  \thispagestyle{empty}

\begin{abstract}
A strong motivation of charging depleted battery can be an enabler for network capacity increase. In this light we propose a spatial attraction cellular network (SAN) consisting of macro cells overlaid with small cell base stations that wirelessly charge user batteries. Such a network makes battery depleting users move toward the vicinity of small cell base stations. With a fine adjustment of charging power, this user spatial attraction (SA) improves in spectral efficiency as well as load balancing. We jointly optimize both enhancements thanks to SA, and derive the corresponding optimal charging power in a closed form by using a stochastic geometric approach.\end{abstract}


%
\IEEEpeerreviewmaketitle

\section{Introduction}

With the mobile data traffic explosion, the small cell network has come to the fore to offload over-utilized macro cell traffic. However, a small cell base station (BS) frequently serves a limited number of users due to its small coverage area  \cite{Nomor}, \cite{Moon}. In addition, the mobility of users even make the time spent in small cell area relatively short. This results in under-utilization of small cell resources \cite{Moon}, motivating us to balance traffic between macro and small cells.

As a remedy for this, we suggest to enable small cell BSs to charge mobile devices wirelessly, given strong desire of mobile users to charge their devices \cite{JD_power}. Seeking for charging depleted batteries even makes users move to a nearby place providing power outlets \cite{intel, star}. Therefore, a wireless power transfer (WPT) in small cell BSs can act as an incentive to yield the spatial attraction (SA) of users in macro cells toward the vicinity of small cell BSs. Throughout this paper, we call this network a spatial attraction cellular network (SAN) and its attraction amount is determined by the charging power. 

The physical displacement from macro to small cells can be utilized for the improvement in macro-to-small cell \emph{load balancing}. From the small cell users perspective who were already in the small cell coverage (see Fig. 1), this attraction also leads to the improvement in \emph{spectral efficiency}. This is because the attraction toward their associated BSs not only increases their received powers of the desired signals but also coincidentally decreases interference.

In an SAN, these load balanging and spectral efficiency improvements vary with charging power, but their corresponding behaviors are non-trivial. From the load balanging perspective, increasing charging power for instance may incur too much associations at small cell BSs. To mitigate such an issue, it requires an optimal charging power that balances the associations between macro and small cell BSs. From the spectral efficiency perspective, on the contrary, high charging power increases the SA users' distances from the associated small cell BSs, at which the charged powers are still satisfactory in that distance (compare (a) and (b) in Fig. 1). As a result, increasing charging power decreases spectral efficiency improvement. In this paper we thus provide the SAN rate maximizing charging power that jointly optimize the load balanging and spectral efficiency improvements.

\begin{figure}     
\centering
   \subfigure[High charging power (large $r_c$)]{\centering
     \includegraphics[width=4.2cm]{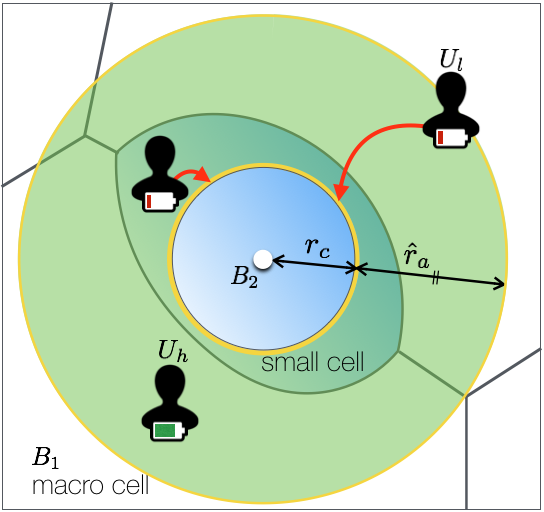} }  \hskip -10pt
   \subfigure[Low charging power (small $r_c$)]{
     \includegraphics[width=4.2cm]{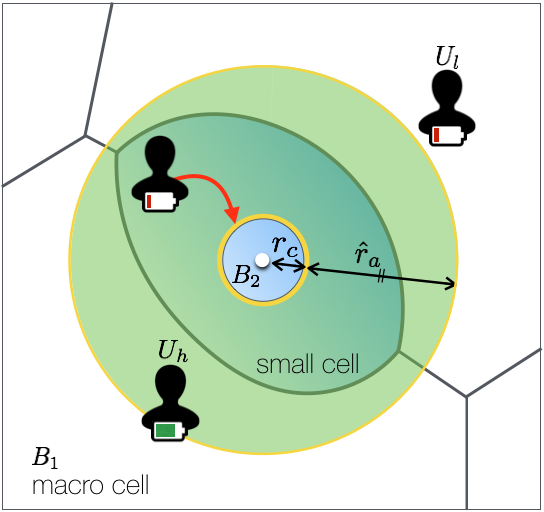}} 
\caption{Illustration of an SAN for different charging powers (or charging range $r_c$) with the maximum SA distance $\hat{r}_a$. A small cell BS $B_2$ spatially attracts battery depleting users $U_\ell$ to its charging rim (yellow thick circle).}
\end{figure}

\textbf{Feasibility of SAN} -- 
SAN implementation hinges on whether SA is viable in practice. Although mobile device proliferation is relentless, improving battery capacity has still been a major technical bottleneck \cite{paradiso}. It makes users more prone to be in danger of battery depletion, leading to a strong battery charging motivation in reality \cite{JD_power}. Such a motivation results in SA toward charging places, for example, when determining where to go for coffee \cite{star, mcdonald} or even when charging is the sole reason for movements \cite{intel}. 

Another key enabler is to implement WPT capable small cells. By means of the advanced signal processing technologies such as an adaptive beamforming, microwave WPT has a potential to provide battery charging for most of small cell coverages while guaranteeing user safety \cite{KHuang:CuttingLast}. When utilizing a beamforming antenna having an aperture of $3$ m in radius, for instance, it theoretically yields charging at the distance of $10$ meters with the charging amount of $0.5$ Watt when the transmit power is $50$ Watt \cite{KHuang:CuttingLast}.

\textbf{Related Works} -- 
WPT aided cellular networks have recently attracted attention, accompanied by rapid advancement in WPT technologies \cite{Kurs_10, Cota}. Its advantage in network capacity perspective has been explored in \cite{Huang_14,swko}. The network capacity enhancement comes from allowing more communicating users in that WPT mitigates battery depletion. Along with such improvement, we also capture the network capacity increment via increasing spectral efficiency owing to SA. The SA also provides improvement via macro-to-small cell load balancing without degrading spectral efficiency. It is unlike preceding schemes such as cell range expansion (CRE) \cite{qualcomm, offloading} that balance the load in return for the spectral efficiency decrement. In detail, offloaded users in CRE are associated with small BSs who give lower signal power than the originally associated macro BSs. Than the originally associated BSs turn into dominant interferers, whose powers are even higher than the recived signal powers, leading to the decrement in spectral efficiency.

\textbf{Contributions} --
This study examines the ramifications of WPT in downlink small cells in terms of maximizing rate-coverage, the proability of a typical user achieving data rate greater than a threshold. The main contributions are listed below.
\begin{itemize}
\item We propose an SAN that spatially attracts users to small cell BSs by the aid of battery charging WPT and thereby elevates the rate coverage of the network. We use a stochastic geometric approach to derive its rate coverage and demonstrate that significant improvements are feasible only with short SA distance, e.g. 125 \% rate coverage when maximum SA distance is only $1$ m (see Propositions 1 and 2 as well as Fig. 3 in Section III).
\item An optimal charging power is derived in a closed form in an asymptotic case where user density is much higher than that of BSs such as an urban area, providing a design guideline of SAN (see Corollary 1 in Section III). 
\item To analyze SAN rate coverate, we consider data usage pattern dependent on user's residual energy. The analytic result shows that the more users' downlink (DL) usages are affected by their residual energies, the more rate coverage SAN can achieve (see Fig. 4).
\end{itemize}


\vskip 10pt \noindent 

\section{SAN System Model} 

\begin{table}[t]
\caption{Summary of Notations} \centering
	\begin{tabular}{c p{6cm}  p{6cm} }
	\hline\hline Notation & Meaning\\[1ex]
	\hline 
	$U_\ell$ (or $U_h$) & Low (or high) battery level user\\
	$B_k$ & $k$-th tier base station for $k \in \{1, 2 \}$\\
	$\Phi_k$ $\(\text{or }\Phi_u\)$ & $B_k$ (or user) locations \\
	$\lambda_k$ $\(\text{or } \lambda_u\)$ & $B_k$ (or user) density \\
	$P_k$ & $B_k$'s information transmission power\\
	$P_{\text{tc}}$ $\(\text{or }P_{\text{rc}} \)$ & Charging transmission (or reception) power\\
	$P_u$ & User energy consumption during a unit time slot\\
	$r_c$ & Maximum charging range to charge $P_u$ \\
	$r_a$ (or $\hat{r}_a$) & User spatial attraction (or maximum) distance \\
	$\mathcal R_k^{\ell}$ $\(\text{or }\mathcal R_k^{h}\)$ & Rate coverage of a low (or high) battery user when associated with $k$-th tier base station \\
	$\mathcal R$ & Average rate coverage of user
	\\ [1ex]
	 \hline
\end{tabular}
\label{table:notation}

\end{table}

\subsection{Information Transmission}
Consider a two-tier downlink cellular network comprising macro BSs $B_1$ and small BSs $B_2$. Let the subscript $k \in \{ 1, 2\}$ hereafter denote the corresponding tier. The $k$-th tier BS coordinates $\Phi_k$ follow a homogeneous Poisson point process (HPPP) with density $\lambda_k$ where $\Phi_1$ and $\Phi_2$ are independent of each other. User locations $\Phi_\text{u}$ also follow a HPPP with density $\lambda_\text{u}$, independent of $\Phi_1$ and $\Phi_2$. For simplicity without loss of generality, we consider a unity time slot between the realizations of the spatial locations. All notations are summarized in Table I. 

A BS in the $k$-th tier $B_k$ transmits a downlink information signal with power $P_k$. The transmitted signal experiences path loss attenuation with the exponent $\alpha >2$ and Rayleigh fading with unity mean. Both $B_1$ and $B_2$ utilize the same frequency, of the bandwidth $W$, yielding in inter-tier interference. Consider an interference-limited environment. By using Slyvnak's theorem \cite{kendall}, the signal-to-interference ratio ($\SIR$) at a typical user associated with a $B_k$ and located at the origin is given as
\begin{align}
\SIR_k:=\frac{P_k  h_k^{(0)} \big|B_{k}^{(0)}\big|^{-\alpha}}{\sum_{j=1}^{2}{\sum_{B_{j}^{(i)} \in \Phi_j} {P_j h_j^{(i)}  \big|B_{j}^{(i)} \big|^{-\alpha} }}}
\end{align}\normalsize
where $B_{k}^{(0)}$ denotes the coordinates of the BS associated with the typical user, $B_{k}^{(i)}$ for $i\in\{1, 2,\cdots\}$ the $i$-th nearest interfering BS coordinates from the typical user, and $h_k^{(0)}$ and $h_k^{(i)}$ the corresponding channel fading gains independently following exponential distribution with unity mean. The notation $|\cdot|$ indicates the Euclidian norm. Multiple users in a BS are served according to a uniformly random scheduler \cite{tse}.

\subsection{User Energy Consumption}
Users download information with probability $u$. Each information reception consumes energy amount $P_u$ per unity time slot. For a user with residual energy amount $L$, it allows at most $i$ times consecutive information receptions when $i P_u \leq L < (i+1) P_u$ for an integer $i\geq 0$. We discretize  $L$ into $L_i$'s according to the maximum number of information receptions $i$ in a way that $i=\lfloor {\frac{L}{P_u}} \rfloor$ where $\lfloor \cdot \rfloor$ is a floor function. For more clarity without loss of generality, we consider battery capacity is $3 P_u$, i.e. $L \in \{L_0, L_1, L_2, L_3 \}$.


Battery depleting users are assumed to attempt less frequent information receptions. This leads to specify the information download probability $u$ as 
\begin{align} \label{eta}
\hspace{10pt} u = \left\{\begin{array}{lll} 
             0 && \text{if $L \in L_0$ (Empty)} \vspace{3pt}\\ 
             u_\ell && \text{if $L \in L_1$ (Low)} \vspace{3pt}\\
             u_h  && \text{if $L \in \{{L_2, L_3}\}$ (High)} \vspace{3pt}
            \end{array}\right.
\end{align}
where $u_h\geq u_\ell \geq 0$. For notational brevity, let $U_\ell$ denotes battery depleting users whose residual battery energy $L \in L_1$. Similarly, $U_h$ indicates the users having $L \in \{ L_2, L_3 \}$.


\subsection{WPT Charging}
Small cell BS $B_2$'s are equipped with additional beamforming antennas dedicated for charging battery via microwave WPT. For charging battery, each user has an antenna which is associated with one of the predefined antenna patterns so that the main lobe point heads for her. Each $B_2$ transmits a charging signal with power $P_{\text{tc}}$ through a power transfer dedicated frequency whose bandwidth is assumed to be unity without loss of generality. The charging channel is separated, yielding no mutual interference. The transmitted charging signal experiences path loss attenuation with the exponent $\beta>2$. Users associate with the nearest $B_2$. The received charging power is
\begin{equation}
P_{\text{rc}}:=
G_m P_{\text{tc}}  \max\left\{ \big| B_2^{(0)} \big|,1 \right\}^{-\beta},
\end{equation}
 where $G_m$ denotes the main lobe gain. By means of highly sophisticated beamforming technique, the charging power received from non-associated $B_2$'s is assumed to be negligibly small. For more brevity, we henceforth consider $P_{\text{tc}}$ is normalized by $P_u$ and always bigger than $P_u$. Since residual battery level is elevated when $P_{\text{rc}} \geq P_u$, the maximum charging range $r_c$ becomes $({P_{\text{tc}}}/{P_u})^{1/\beta}$ which makes $P_{\text{rc}}$ be equivalent to $P_u$.

\subsection{Spatial Attraction}
All users tend to keep their positions, but the battery depleting users are willing to move for charging a unit energy amout $P_u$ with the SA distance $r_a (\leq \hat{r}_a)$. Specifically, assuming $B_{2}$'s broadcast their locations, a user is spatially attracted toward its nearest $B_2$ if the following two SA conditions hold: (i) $L \in \{L_0, L_1\}$ and (ii) $r_a \leq \hat{r}_a$. The former indicates users are battery depleting (or $U_\ell$), and the latter represents SA distance should be no greater than the users' maximum feasible SA distance. The required amount of energy to receive the $B_2$ location information is assumed to be negligible; so that even $L_0$ users can successfully receive it. Such SA affects user locations, and thus results in user mobility model specified by the following two phases. For each unity time slot, they occur in order.

\begin{enumerate}[$\hspace{10pt}1.$]
\item \emph{(Uniform Distribution):} Users are uniformly distributed.
\item \emph{(Spatial Attraction):} SA users move toward their nearest $B_2$ through the shortest paths until reaching $B_2$'s charging rim providing $P_u$ (see yellow thick circle in Fig. 1); non-SA users keep their locations.
\end{enumerate}

\vskip 10pt \noindent 

\section{SAN Rate Coverage Maximization}

\begin{figure}
\centering 
{\includegraphics[width=8.8cm]{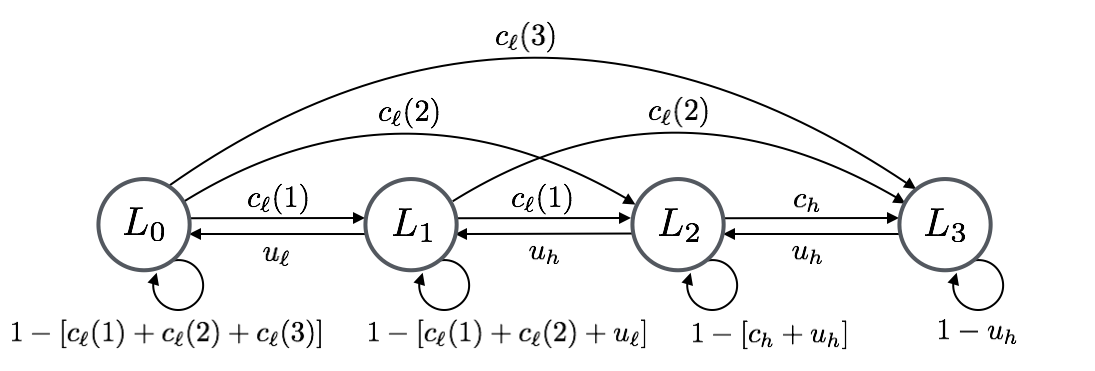}} 
\caption{Four-level residual battery energy Markov chain where the state $L_i$ corresponds to the maximum $i$ number of information receptions.} \label{Fig3}
\end{figure}

In this section we focus on WPT charging power $P_{\text{tc}}$, as well as its corresponding $r_c$, that maximizes average rate coverage $\mathcal{R}$ in an SAN, the probability that a typical user's average downlink rate exceeds a target threshold $\theta$. We rigorously specify the definition as
\begin{align}
\mathcal R := \E_{\Phi_k}\left[ \frac{u}{N_k+1}  \mathbb{P}\big( \log [1+\textsf{SIR}_k]>\frac{\theta}{W}\big) \right]
\end{align} \normalsize
where $N_{k}$ represents the number of users in a $B_{k}$ associated with a typical user.

\begin{figure*}[b]
\hrulefill
{\small
\begin{align} \nonumber
\mathcal R_k^{h} & =    \frac{1}{1+\rho_{(\theta, M)}} \\  \nonumber
\mathcal {R}_1^{\ell} &= {\frac{e^{-\pi\lambda_{2}{\({P_{tc}}^{\frac{1}{\beta}}+\hat r_a \)}^{2} \text L_2(\theta) }}{{\text L_1(\theta)}{\mathcal A^{\ell}_1}}} \  + \ \frac{e^{-\pi\lambda_{2}{\({P_{tc}}^{\frac{1}{\beta}}+\hat r_a \)}^{2}}}{{\mathcal A^{\ell}_1}} \int_{0}^{M}{r e^{-\pi\lambda_{1}r^2 \left[1+ \rho_{(\theta, M)}+\lambda \rho_{(\theta,r)}\right] }} dr  \\ \nonumber
\mathcal R_2^{\ell}  &= {\frac{{e^{-\pi\lambda_{2}{\({P_{tc}}^{\frac{1}{\beta}}+\hat r_a \)}^{2}\text{L}_2(\theta) }}}{{{\text{L}_2(\theta)}^{-1}\mathcal A^{\ell}_2}}+\frac{{{\mathcal A_2^{h}}-{\mathcal A_2^{h}} e^{-\pi\lambda_{2}{{P_{tc}}^{\frac{2}{\beta}}}\left[1+\frac{\rho_{(\theta, M)}}{\mathcal A_2^{h}}\right]} }}{{\mathcal A^{\ell}_2}{\mathcal A_2^{h}}+{{\mathcal A^{\ell}_2}}\rho_{(\theta, M)}} \ + \ \frac{e^{-\pi \lambda_2{{P_{tc}}^{\frac{2}{\beta}}} \left[1+\frac{\rho_{(\theta, M)}}{\mathcal A_2^{h}}\right]}- e^{-\pi \lambda_2{{P_{tc}}^{\frac{2}{\beta}}} \left[\frac{\rho_{(\theta, M)}}{\mathcal A_2^{h}} \ + \ \left(1+\frac{\hat r_a }{{P_{tc}}^{\frac{1}{\beta}}}\right)^2\right]}}{\mathcal A_2^{\ell}}} \quad\quad  \end{align}
\small\begin{align}
 \rho_{(\theta,x)}={(e^{\frac{\theta}{W}}-1)}^{\frac{2}{\alpha}} \int_{\left(\frac{M}{x}\right)^2(e^{\frac{\theta}{W}}-1)^{-\frac{2}{\alpha}}}^{\infty}{{\left(1+u^{\frac{\alpha}{2}}\right)}^{-1}} du , 
\text L_k(\theta)=\tfrac{1+\rho_{(\theta, M)}}{\mathcal A^h_k}, 
M=\({P_{tc}}^{\frac{1}{\beta}}+\hat r_a \)\left(\tfrac{P_1}{P_2}\right)^{\frac{1}{\alpha}}, 
\lambda=\frac{\lambda_2}{\lambda_1}\left(\tfrac{P_2}{P_1} \right)^{\frac{2}{\alpha}} \quad \label{Eq:RateCov1}
\end{align}} 
\end{figure*}

\subsection{Association Probability}

In SAN, a user is associated with a BS based on its residual energy and the received power strength. 

\vskip 10pt \noindent\begin{lemma} \emph{The probability $\mathcal A_k^h$ (or $\mathcal A_k^\ell$) that $U_h$ (or $U_\ell)$ associates with $B_k$ is given as
\begin{align}
\mathcal{A}^h_k&=\[{{\lambda_k}^{-1}\sum_{i=1}^2 \lambda_i {\(\frac{P_i}{P_k}\)}^{\frac{2}{\alpha}}}\]^{-1}, \\ 
\mathcal{A}^\ell_k&=(-1)^k\(\mathcal A^h_2 e^{-\pi \tfrac{\lambda_2}{\mathcal A^h_2}{r_s}^2  }-e^{-\pi\lambda_2 {r_s}^2}\)+(k-1)
\end{align}\normalsize where $r_s=r_c+\hat r_a$.  \vskip 10pt \noindent{\bf
Proof.}  See Appendix-A.
 \hfill $\blacksquare$}
\end{lemma}

Such a BS association changes the association distance as well as the number of associated users. They are derived by utilizing Lemma 1, to produce the rate coverage.


\subsection{Rate Coverage and Its Optimal Charging Power}

To state our main result, the rate coverage, let $\mathbf{q}$ denote the steady-state probability vector of each residual battery energy level, $\mathbf q = [q_0, q_1, q_2, q_3]$, whose states and transition probabilities are shown as a Markov chain in Fig. 2. SAN rate coverage $\mathcal{R}$ is represented in the following  proposition.

\vskip 10pt \noindent\begin{proposition}\emph{
Rate coverage in an SAN is given as
\small\begin{align}
\mathcal R=\sum_{k=1}^2\left(\sum_{n\geq1}\frac{\mathbb{P}(N_{k}=n)}{n+1}\left[{ {P_H}\mathcal A_k^h\mathcal R_k^{h} + {P_L} \mathcal A_k^{\ell}\mathcal R_k^{\ell}}\right]\right).
\end{align}\normalsize
The residual battery energy dependent information reception probabilities $P_H = (q_2+q_3) u_h$ and $P_{L}= q_1 u_\ell$ for $q_{1}$, $q_{2}$, and $q_{3}$, which are obtained by solving $\mathbf q\mathbf{T}=\mathbf{q}$ where
\setlength\arraycolsep{0em}
 \medmuskip=0mu\thinmuskip=0mu\thickmuskip=0mu 
\begin{equation}
\mathbf{T}=\[ 
\begin{array}{cccc}
1- \sum_{i=1}^3c_\ell(i) & c_\ell(1) & c_\ell(2) & c_\ell(3) \\ 
u_\ell & 1-u_\ell-  \sum_{i=1}^2c_\ell(i) & c_\ell(1) & c_\ell(2) \\
0 & u_\ell & 1-u_\ell-c_\ell(1) & c_\ell(1) \\
0 & 0 & u_h & 1-u_h 
\end{array}\], \nn
\end{equation}
\begin{align} \nonumber
c_\ell(1)&=e^{-\pi\lambda_2 2^{-\frac{2}{\beta}} {r_c}^2}-e^{\pi\lambda_2 {(r_c+\hat r_a)}^2},  \\ \nonumber
c_\ell(2)&=e^{-\pi\lambda_2 3^{-\frac{2}{\beta}} {r_c}^2}-e^{-\pi\lambda_2 2^{-\frac{2}{\beta}} {r_c}^2}, \\ \nonumber
c_\ell(3)&=1-e^{-\pi\lambda_2 3^{-\frac{2}{\beta}} {r_c}^2},\text{ and}\\ \nonumber
c_h&=1-e^{-\pi\lambda_2 {r_c}^2}.
\end{align}
\\ The distribution of $B_k$ associated user number is $\mathbb{P}\(N_k=n\)=\sum\limits_{m=0}^n\mathbb{P}\(N_k^{h}=n-m\)\mathbb{P}\(N_{k}^{\ell}=m\)$ where
\medmuskip=-3mu\thinmuskip=-3mu\thickmuskip=-3mu
{\small\begin{align}
\mathbb{P}\(N_k^{h}=n\)&=\frac{3.5^{3.5}\Gamma(n+4.5)}{n!\Gamma(3.5)}\left(\frac{\lambda_uP_H\mathcal A_k^{h}}{\lambda_k}\right)^n\left(3.5+\frac{\lambda_uP_H\mathcal A_k^{h}}{\lambda_k}\right)^{-(n+4.5)} \text{,} \nn \\
\mathbb{P}\(N_k^{\ell}=n\)&=\frac{3.5^{3.5}\Gamma(n+4.5)}{n!\Gamma(3.5)}\left(\frac{\lambda_uP_L\mathcal A_k^{\ell}}{\lambda_k}\right)^n\left(3.5+\frac{\lambda_uP_L\mathcal A_k^{\ell}}{\lambda_k}\right)^{-(n+4.5)}. \nn
\end{align}}
\noindent Given residual battery energy and associations, rate coverages are in \eqref{Eq:RateCov1} at the bottom of the page.\vskip 10pt
\noindent{\bf Proof.}  
See Appendix-B. \hfill $\blacksquare$}
\end{proposition}
\medmuskip=1mu\thinmuskip=1mu\thickmuskip=1mu
\noindent

It is worth noting that SAN rate coverage does not monotonically increase with the charging power $P_{\text{tc}}$ as shown in Fig. 3(b). The reason is explained by the following two perspectives.
\begin{enumerate}[$\hspace{10pt}1)$]
\item \emph{The trade-off between load balanging and spectral efficiency gains:} Increasing $P_{\text{tc}}$ (or $r_c$) attracts more users toward small cell areas for load balanging while reducing the spectral efficiency improvement that can be achieved by shortening the distance to BS.
\item \emph{SA motivation:} Too small power $P_{\text{tc}}$ motivates few SA that cannot sufficiently operate an SAN so as to provide a rate coverage improvement. Charging too much power, in contrast, retains mobile users' residual energy enough and may lose their SA motivations, decreasing the improvements in both load balanging and spectral efficiency.
\end{enumerate}

This makes us turn our attention to derive the optimal charging power $P_{\text{tc}}^*$ maximizing rate coverage. This derivation is not straightforward due to the user safety condition. The received charging power density of a user should not exceed the maximum safe power density $\eta$ \cite{safety}, delimiting the feasible range of $P_{\text{tc}}^*$. Furthermore, the maximum feasible amount of $P_{\text{tc}}^*$, as well as its corresponding distance $r_{c}$, is also restricted by $B_{2}$ Voronoi cell inscribed circle radius $\nu$; otherwise, SA does not always increase spectral efficiency and load balancing, which may degrade rate coverage. We collectively consider such trade-off and constrains on $P_{\text{tc}}$, and yield its optimal value.

\begin{figure*}     
\centering
   \subfigure[For $\lambda_u = 2 \times 10^{4}$ users/$\text{km}^{2}$, $\lambda_2 = 3 \times 10^{2} \text{ BSs}/\text{km}^2$]{\centering
     \includegraphics[width=8.8cm]{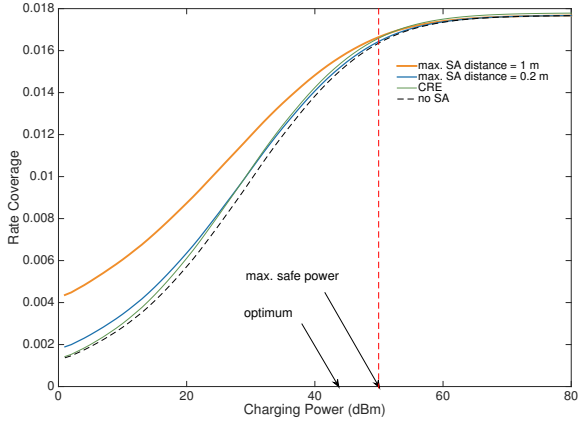} }
   \subfigure[For $\lambda_u = 1  \times 10^{7}$ users/$\text{km}^{2}$, $\lambda_2 = 3 \times 10^{3} \text{ BSs}/\text{km}^2$]{
     \includegraphics[width=8.8cm]{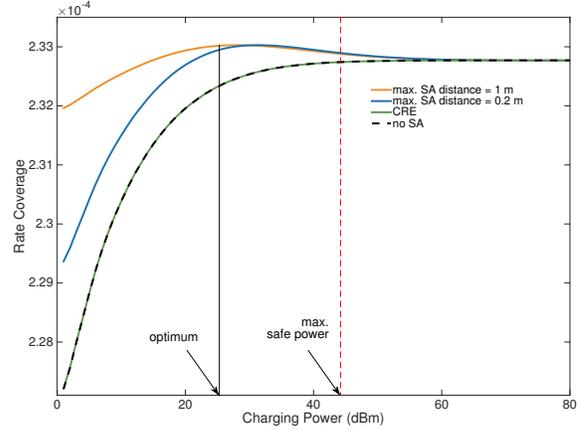}} 
\caption{Rate coverage in an SAN ($\lambda_1= 10\text{ BSs}/\text{km}^2$, $P_1=43$ dBm, $P_2=23$ dBm, $W=10$ MHz, $\theta=1$ Mbps, $\alpha=4$, $\beta=5$, $\eta=10$ $W/m^2$, $G_m=20$ dBi, $G_s=-6.5$ dBi \cite{Ram_01}).}
\end{figure*}
 
\vskip 10pt \noindent\begin{proposition}\emph{Assuming the load at each BS is equal to its mean to simplify SAN rate coverage, the optimal charging power $P_{\text{tc}}^*$ is given as
{\begin{align}
P_{\text{tc}}^*=\min\left\{\text{arg}\underset{P_{\text{tc}}}{\operatorname{max}}\sum_{k=1}^2\frac{{ {P_H}\mathcal A_k^h\mathcal R_k^{h} + {P_L} \mathcal A_k^{\ell}\mathcal R_k^{\ell}}}{1+1.28\tfrac{\lambda_u}{\lambda_k}\( {P_H}\mathcal A_k^h+ {P_L}\mathcal A_k^{\ell} \)},\hat P_{\text{tc}}^{\eta},\hat P_{\text{tc}}^{\nu}\right\}
\end{align}}
where 
{\begin{align} \nonumber
\hat P_{\text{tc}}^{\eta}&= \frac{2 G_m \lambda_2 P_{\text{tc}}}{\tan^2(\frac{t_m}{2})} \[1-e^{-\lambda_2 \pi} - \lambda_2 \pi  \text{Ei}\(-\lambda_2 \pi \)  \] ,\text{ and}\\
\hat P_{\text{tc}}^{\nu}&=\frac{1}{G_m} \left({ 16\lambda_2+4\left[1+{\left(\frac{P_1}{P_2}\right)}^{\frac{1}{\alpha}}\right]^2\lambda_1 }\right)^{-\frac{\beta}{2}}
\end{align}
} \normalsize
where $t_m$ denotes a beam width and $\text{Ei}(x)$ an exponential integral function.
\vskip 10pt \noindent
{\bf  Proof.} 
Appendix-C
 \hfill $\blacksquare$
}
\end{proposition}
Although an optimal charging power $P_{\text{tc}}^*$ derivation is not possible in a closed form, it can be discovered through a linear search utilizing the above proposition, which is easily accessible compared to running an exhaustive simulation. Furthermore, to get a closed form optimal value, we consider an asymptotic case where the user density is much higher than BSs'.

\vskip 10pt \noindent\begin{corollary}\emph{For $\lambda_u\gg\lambda_k$, $P_{\text{tc}}^*$ is given as below.
\begin{itemize}
\item\emph{When $\lambda_1 \ll \lambda_2$,}
\small\begin{align}\small
P_{\text{tc}}^*\ \approx \  \frac{1}{G_m}\left[\left(\frac{1+{\rho_{(\theta,T)}}}{6\pi \[\lambda_1\left(\frac{P_1}{P_2}\right)^{\frac{2}{\alpha}}+\lambda_2\]}-\frac{{\hat r_a}^2}{12}\right)^\frac{1}{2}-\frac{1}{2}\hat r_a\right]^{\beta}
\end{align}\normalsize
\item\emph{When $\lambda_1 \geq \lambda_2$,}
 \small\begin{align}\small
P_{\text{tc}}^*\ \approx\  \frac{1}{G_m} \left[\left(27\pi\rho_{(\theta,T)}\left[\lambda_1\left(\frac{P_1}{P_2}\right)^{\frac{2}{\alpha}}+\lambda_2\right]\right)^{-\frac{1}{2}}-\frac{3}{2}\hat r_a\right]^{\beta}.
\end{align}\normalsize
\end{itemize}\vskip 10pt
\noindent
{\bf  Proof.} 
Appendix-D
 \hfill $\blacksquare$
}
\end{corollary}

\normalsize

The result provides a charging power control guideline in an SAN as follows. For low small cell density $\lambda_2$ and/or the SA distance $\hat r_a$, the SAN is likely to suffer from macro cell traffic congestion, requiring to attract more users to small cells for load balanging. In such a scenario, $P_{\text{tc}}^{*}$ should be increased, and vice versa for the opposite situation. For a large threshold $\theta$, spectral efficiency dominates the rate coverage, and thus $P_{\text{tc}}^{*}$ should be decreased to shorten the association distances, increasing spectral efficiency.


Fig. 3 illustrates the SAN rate coverage that is superior to the case with no SA, and the improvement increases along with SA motivation (or maximum SA distance $\hat{r}_a$). Moreover, it captures the proposed network outperforms a network with CRE that is beneficial to load balancing yet is harmful to spectral efficiency \cite{qualcomm, offloading}. Specifically, when it comes to load balancing, CRE resorts to decreasing the received power from an associated BS. In an SAN, on the contrary, its load balancing coincidentally shortens user association distance, and therefore also increases the received power from an associated BS, leading to the further improvement in rate coverage than CRE. It is worth mentioning that if users are willing to move at least $20$ cm for battery charging, an SAN become superior in rate coverage to the network with CRE. Though, SAN may achieve lower rate coverage than CRE if users are unlikely to be attracted, as shown in Fig. 3(a).

Regarding a user safety requirement, the optimal charging power may exceed the maximum safe charging power (red dotted lines). However, such a situation can be detoured via high user density or deploying more small cells as in Fig. 3(b).

Fig. 4 depicts the maximized SAN rate coverage monotonically increases with small cell BS density. This relentless improvement is expected to be different from the case with CRE whose improvement is saturated for high small cell BS density where load balancing is not much needed (see the convergence of dotted black and solid green lines in Fig. 3(b)). SAN can increase rate coverage via its spectral efficiency improvement even viable in such an environment. Focusing on the exponentially decreasing optimal charging power along with small cell BS density, deploying small cells can be a viable solution for ever-growing rate coverage while abiding by a safety requirement.

In addition, the proposed SAN can improve much more rate coverage when DL traffic usage ratio of low-to-high battery users is large, i.e. when users are more sensitive to their residual energy. This represents that SAN effects well along with a growth of booming wireless communication services that require large battery consumptions.

 \begin{figure}
\centering
\includegraphics[width=8.8cm]{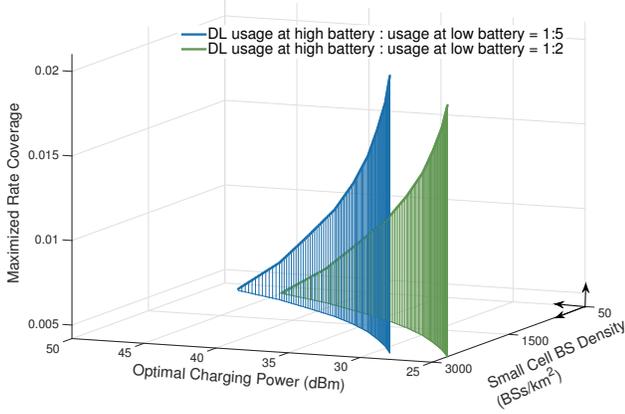} 
\caption{Maximized rate coverage and optimal charging power with respect to the $\lambda_2$ and DL usage ratio $\frac{u_h}{u_{\ell}}$ ($\lambda_1 = 10 \text{ BSs}/\text{km}^2$, $\lambda_{u} = 1 \times 10^{4}$ users/$\text{km}^{2}$, $P_1=43$ dBm, $P_2=23$ dBm, $W=10$ MHz, $\theta=1$ Mbps, $\alpha=4$, $\beta=5$, $\hat{r}_a=2 \text{ m}$, $G_m=10$ dBi, $G_s=-7.4$ dBi \cite{Ram_01}).}
\end{figure} 

\vskip 10pt \noindent 

 \section{Conclusion}

In this paper we have analyzed an SAN that spatially attracts battery depleting users by means of providing WPT battery charging. The result sheds light on WPT charging power control in order to maximize rate coverage of the SAN. Our analytic result reveals that charging power increase as much as possible does not guarantees the maximum rate coverage, and thus its optimization is required. Such an optimal charging power is derived in a closed form, and its corresponding design guideline is presented for different network deployment, wireless channel environment, and a user safety requirement.

The weakness of this study is its one-sided focusing on a downlink scenario. SA has a potential to further improve uplink rate coverage by its shrinking association distances. In such a scenario, the uplink optimal charging power may not be in accordance with the downlink optimal charging power. Further extension to this work therefore should involve their joint optimization problem. Another interesting avenue for future work is a network economic analysis on the SAN. In a similar way to \cite{jhpark}, the SAN rate coverage optimization will jointly incorporate BS density and spectrum amount for communications, also combined with such values for WPT, providing a network operator's SAN deployment guideline.

\vskip 10pt \noindent 

\section*{Appendix}

\subsection{Proof of Lemma 1}
For $U_{h}$, the association probability $\mathcal A_{k}^{h}$ is directly derived by using Lemma 1 in \cite{fca}. For $U_\ell$, the association probability $\mathcal A_k^\ell$ should incorporate the SA impact. Let $r_k$ denote the distance between a typical $U_\ell$ and his nearest $B_k$. Now that a $U_\ell$ is attracted toward the nearest $B_2$ if $r_{2} < r_s$, $\mathcal A_1^\ell$ is given as
\begin{align}\small \nonumber
\mathcal A_1^\ell
&= \mathbb{P} (U_\ell \rightarrow B_1 \text{ and } r_2 \geq r_s)\\ \nonumber 
&=\mathbb{P} \left( P_1 {r_1}^{-\alpha} \geq P_2 {r_2}^{-\alpha} \text{ and } r_2 \geq r_s \right) \\ \nonumber &\overset{(a)}{=} 2\pi\lambda_1  \displaystyle \int_{r_s\left(\frac{P_1}{P_2}\right)^{\frac{1}{\alpha}}}^\infty r e^{-\pi r^2 \sum_{i=1}^2 \lambda_i \left({\frac{P_i}{P_1}}\right)^{\frac{2}{\alpha}}} dr
\\ &\ \ \ +  2\pi\lambda_1 e^{-\pi\lambda_2{r_s}^2}\int_0^{{r_s\left(\frac{P_1}{P_2}\right)^{\frac{1}{\alpha}}}}re^{-\pi\lambda_1r^2} dr
\end{align}\normalsize
where (a) follows from the nearest neighbor distance distribution \cite{kendall}. Similarly, $\mathcal A_2^\ell$ is given as
\begin{align}  
\mathcal A_2^\ell  \nonumber
&= \mathbb{P}(U_\ell \rightarrow B_{2}\text{ or } r_2 < r_s)\\  \nonumber
&= \mathbb{P}\left(P_2 {r_2}^{-\alpha} \leq P_1 {r_1}^{-\alpha}, \text{ } r_2 \geq r_s \right) + \mathbb{P}(r_2<r_s) \nonumber \\ &= \displaystyle 2\pi\lambda_2\int_{r_s}^\infty r e^{-\pi\left(\sum_{i=2}^2 \lambda_i\frac{P_i}{P_2}^{2/\alpha}+\lambda_2 \right)r^2}dr+1-e^{-\pi\lambda_2{r_s}^2}  \label{ap2}
\end{align}\normalsize
that finalizes the proof.
 \hfill $\blacksquare$

\subsection{Proof of Proposition 1}
The desired result comes from the following four parts.
\subsubsection{Residual Energy Dependent Information Reception Probabilities $P_{H}$ and $P_{L}$}
Let $c_{{\ell}}(i)$ denote the probability that $U_{\ell}$ is capable of additional $i$ times information receptions thanks to WPT battery charging. Its calculation is followed by the fact that $U_{\ell}$ can receive $i$ times information receptions thanks to battery charging if $2^{-\frac{2}{\beta}}r_c<r_{2} \leq r_{c}+\hat r_{a}$ for $i=1$ and if $(i+2)^{-\frac{1}{\beta}}r_c<r_{2}\leq (i+1)^{-\frac{1}{\beta}}r_c$ for $i=2, 3$. Let $c_{h}$ denote the battery charging probability of $U_{h}$. The event of $c_{h}$ is identical to the event that $U_{h}$ is located with the distance closer than $r_{c}$ from the nearest $B_{2}$. Applying void probabilities \cite{kendall} to their spatial regions of $c_{\ell}(i)$ and $c_{h}$ yield the calculations.

\subsubsection{BS Association Distance Distribution}
Now that an SA user goes toward the nearest $B_2$, the distance between them diminishes. For the SA affected association distance for $U_{\ell}$, its complementary cumulative distribution function (CCDF) is 
\begin{align}
\mathbb{P}(r_k> r|U_\ell\rightarrow B_k)=\frac{\mathbb{P}(r_k>r \text{ and }U_\ell \rightarrow B_k)}{\mathcal A_k^{\ell}}
\end{align}
where $U \rightarrow B_{k}$ denotes an event that a user $U$ associates with a $B_{k}$. For the probability density function (PDF) $f^{\ell}_{r_1}(r)$, calculating the numerator in CCDF requires a simple modification of \eqref{ap2} in the proof of Lemma 1. For $f^{\ell}_{r_2}(r)$, the numerator calculation is divided into two cases: (i) if $r_c\leq r_2<r_s$, an $U_\ell$ is located at a distance of $r_c$ from $B_2$; (ii) if $r_2<r_c$, $r_{c}$ is restricted to the $B_2$ cell, making the calculation become the void probability \cite{kendall} of a ball with radius $r_{2}$. 
Applying Lemma 1 and differentiating CCDF derive $f^{\ell}_{r_k}(r)$ as below.
\begin{align} 
f^{\ell}_{r_1}(r) &= \begin{cases} 
\frac{2\pi \lambda_1}{\mathcal A_1^{\ell}} r e^{-\pi \sum_{i=1}^2 \lambda_i \left(\frac{P_i}{P_1}\right)^{\frac{2}{\alpha}}r^2} & {\textrm{if $  r < r_s{\left(\frac{P_1}{P_2}\right)}^{\frac{1}{\alpha}}$}} \\
\frac{2\pi \lambda_1}{\mathcal A_1^{\ell}} e^{-\pi\lambda_2 {r_s}^2 }re^{-\pi \lambda_1 r^2}  &{\textrm{{otherwise} }}  
\end{cases}, \\
f^{\ell}_{r_2}(r) &= \begin{cases} 
\frac{2\pi \lambda_2}{\mathcal A_2^{\ell}} r  e^{-\pi\lambda_2 r^2} & {\textrm{if $r<r_c$}} \\
\delta_{r_c}\frac{e^{-\pi \lambda_2 {r_c}^2}-e^{-\pi \lambda_2 {r_s}^2}}{\mathcal A_2^{\ell}}  &{\textrm{{if $ r_c\leq r < r_s $} }}   \\ 
\frac{2\pi \lambda_2}{\mathcal A_2^{\ell}} r e^{-\pi \sum_{i=1}^2 \lambda_i \left(\frac{P_i}{P_2}\right)^{\frac{2}{\alpha}}r^2}  &{\textrm{{otherwise} }}  
\end{cases}
\end{align}\normalsize
where $\delta_{{r_{c}}}$ is a Dirac delta function yielding $1$ for $r=r_c$, otherwise $0$. The PDF $f^h_{r_k}(r)$ for $U_h$ is provided by Lemma 4 in \cite{offloading} as $\frac{2\pi\lambda_k}{\mathcal A_k^{h}}re^{-\pi r^2\sum\limits_{i=1}^2 \lambda_i\left(\frac{P_i}{P_k}\right)^{\frac{2}{\alpha}}}$. 

\subsubsection{Distribution of $B_k$ Associated User Number}
The probability mass function of the user number $N_k$ excluding a typical user associated with a $B_k$ is derived by using Lemma 1 and Corollary 1 in \cite{offloading}. 

\subsubsection{Rate Coverages Conditioned on Residual Energy and an Association $\mathcal R_k^{h}$ and $\mathcal R_k^{\ell}$}
Applying Theorem 1 in \cite{offloading} with the preceding results and Lemma 1 enables to calculate the rate coverages. When it comes to averaging interferer locations, it is worth noting that small cell interferer distances at a $B_{1}$ associated typical $U_{\ell}$ range from $r_{s}$ to $\infty$ if $r_{1}<M$; otherwise, the interferer distances range from $0$ to $\infty$. \hfill $\blacksquare$

\subsection{Proof of Proposition 2}
For analytical tractability, we approximate the number of the $k$-th tier associated users as its average value as below \cite{offloading}.

\begin{align}
\E\[N_k\]=\frac{1.28\lambda_u}{\lambda_k}\(P_H \mathcal A_k^h + P_L \mathcal A_k^\ell\).
\end{align}
The feasible range of charging power is restricted by a safety requirement. Since a user recieves the charging power with main lobe gain from his associated $B_2$ and received power from other $B_2$'s is assumed to be negligibly small, the average received power density $\dot P_{rc}$ is represented as

\begin{align}\nonumber
\dot P_{\text{rc}} &\overset{(a)}{=} \E_{\Phi_2} \[ \frac{G_m P_{\text{tc}}}{\pi \tan^2(\frac{t_m}{2}) {\max \left\{\big| B_2^{(0)} \big|,1\right\} }^2} \] \\ \nonumber 
&= \frac{2 G_m \lambda_2 P_{\text{tc}}}{\tan^2(\frac{t_m}{2})} \( \int_0^1  r e^{-\lambda_2 \pi r^2} dr + \int_1^\infty \frac{ r e^{-\pi\lambda_2 r^2}}{r^2}  dr\)    \\ \nonumber
&= \frac{2 G_m \lambda_2 P_{\text{tc}}}{\tan^2(\frac{t_m}{2})} \[1-e^{-\lambda_2 \pi} - \lambda_2 \pi  \text{Ei}\(-\lambda_2 \pi \)  \] 
\end{align}\normalsize where (a) follows from surface area of power transmission in main lobe \cite{Ram_01}, $t_m$ denotes a beam width, and $\text{Ei}(x)$ an exponential integral function.

Applying the safety requirement $\dot P_{rc} \leq \frac{\eta}{P_u}$ results in $\hat{P}_{tc}^\eta$. Another factor delimiting the feasible range of charging power is Voronoi cell area, i.e. the maximum charging distance $r_c$ is no larger than the inradius of a $B_2$ cell area $\nu$. To simplify our exposition, we approximate $\nu$ as its expected value $\E[\nu]$ calculated by using \cite{calka}. 
\begin{align}
\E[\nu] = ({ 16\lambda_2+4[1+{({P_1}/{P_2})}^{{1}/{\alpha}}]^2\lambda_1 })^{-{1}/{2}}.
\end{align}
Applying the maximum charging range requirement $\hat P_{tc}^{\nu}=\frac{{\E[\nu]}^{{\beta}}}{G_m}$ completes the proof.
\hfill $\blacksquare$ 

\subsection{Proof of Corollary 1}
When $\lambda_u \gg \lambda_k$, $\mathcal{R} \approx {P_L \mathcal A_2^{\ell} (\mathcal R_2^\ell-\mathcal R_2^h)}{(P_H \mathcal A_2^{h}+ P_L\mathcal A_2^{\ell})}^{-1}$. Applying Taylor expansion with a linear interpolation simplifies $\mathcal{R}$ as $\pi\frac{\lambda_2}{\mathcal A_2^h}\rho_{(\theta,T)}\(4{\hat r_a}^3{r_c}+6{\hat r_a}^2{r_c}^2+4{\hat r_a}{r_c}^3+{\hat r_a}\) -\(1+\rho_{(\theta,T)}\)\(2 \hat r_a r_c + {\hat r_a}^2\)$ for $\lambda_1 \ll \lambda_2$. In a similar way, the approximation of $\mathcal{R}$ for $\lambda_1 \geq \lambda_2$ becomes $\pi\lambda_2\(2\hat r_a r_c+{\hat r_a}^2\) -\pi^2 {\lambda_2}^2 \frac{\rho_{(\theta,T)}}{\mathcal A_2^h}\(2\hat r_a {r_c}^3+{\hat r_a}^2 {r_c}^2\) $. Differentiating the approximated $\mathcal{R}$ with respect to $r_{c}$ yields the optimal charging power.
\hfill $\blacksquare$ 

\section*{Acknowledgement}
This work was supported by Institute for Information \& communications Technology Promotion (IITP) grant funded
 by the Korea government (MSIP) (No.B0126-16-1017, Spectrum Sensing and Future Radio Communication Platforms).


\vskip 10pt \noindent 

\begin{thebibliography}{1}

\bibitem{Nomor}
V. Pauli, J. D. Naranjo, and E. Seidel,
\newblock ``Heterogeneous lte networks and inter-cell interference coordination,"
\newblock Nomor Research GmbH,
\newblock December
\newblock 2010.

\bibitem{Moon}
J.-M. Moon, J. Jung, S. Lee, A. Nigam, S. Ryoo, 
\newblock ``On the trade-off between handover failure and small cell utilization in heterogeneous networks," 
\newblock in {\it Proc. IEEE International Conference on Communication Workshop (ICCW),}
\newblock London,
\newblock United Kingdom,
\newblock Jun. 2015



\bibitem{JD_power}
J.D. Power,
\newblock ``Battery life: Is that all there is?,"
\newblock Apr. 2012,
\newblock {\it available at: http://tiny.cc/jdpower}

\bibitem{star}
N. Friedman,
\newblock ``Wireless charging stations means even Starbuck haters will soon be ordering frappuccinos,"
\newblock Jul. 2014, 
\newblock {\it available at: http://tiny.cc/sbwpt}

\bibitem{intel}
Intel,
\newblock ``Intel survey: tech norms for travelers,"
\newblock {\it Intel News Release}, 
\newblock Jun. 2012.

\bibitem{paradiso}
J.A. Paradiso and T. Starnet,
\newblock``Energy scavenging for mobile and wireless electronics,"
\newblock {\it IEEE Pervasive Computing},
\newblock vol. 4,
\newblock no. 1,
\newblock pp. 18--27,
\newblock 2005.


\bibitem{mcdonald}
Alan F.,
\newblock ``McDonald's is bringing Qi wireless charging to selected U.K. restaurants,"
\newblock Jan. 2015,
\newblock {\it available at: http://tiny.cc/mcwpt}


%
%

\bibitem{KHuang:CuttingLast}
K. Huang and X. Zhou,
\newblock ``Cutting last wires for mobile communications by microwave power transfer,"
\newblock {\it IEEE Commun. Mag.,}
\newblock vol. 53,
\newblock no. 6,
\newblock pp. 86--93,
\newblock 2015.



\bibitem{Kurs_10}
A. Kurs, R. Moffatt, and M. Soljacic,
\newblock ``Simultaneous mid-range power transfer to multiple devices,"
\newblock {\it Appl. Phys. Lett.},
\newblock vol. 96,
\newblock no. 4,
\newblock 2010.
%
%


\bibitem{Cota}
TechCrunch,
\newblock ``Cota by Ossia aims to drive a wireless power revolution and change how we think about charging,"
\newblock Sep. 2013,
\newblock {\it available at: http://tiny.cc/tchcrunch}

%


%

\bibitem{swko}
S.-W. Ko, S. M. Yu, and S.-L. Kim.,
\newblock ``The capacity of energy-constrained mobile networks with wireless power transfer,"
\newblock {\it IEEE Commun. Lett.},
\newblock vol. 17,
\newblock no. 3,
\newblock pp. 529--532,
\newblock 2013.

\bibitem{Huang_14}
K. Huang and V. K. N. Lau,
\newblock ``Enable wireless power transfer in cellular network: architecture, modeling and deployment,"
\newblock {\it IEEE Trans. Wireless Commun.},
\newblock vol. 13,
\newblock no. 2,
\newblock pp. 902--912,
\newblock 2014.


\bibitem{qualcomm}
A. Damnjanovic et al.,
\newblock ``A survey on 3GPP heterogeneous networks,"
\newblock {\it IEEE Wireless Commun. Mag.},
\newblock vol. 18,
\newblock no. 3,
\newblock pp. 10--21,
\newblock 2011.


\bibitem{offloading}
S. Singh, H. S. Dhillon, and J.G. Andrews,
\newblock ``Offloading in heterogeneous networks: Modeling, analysis, and design insights,"
\newblock {\it IEEE Trans. Wireless Commun.},
\newblock vol. 12,
\newblock no. 5,
\newblock pp. 2484--2497,
\newblock 2013.


\bibitem{kendall}
D. Stoyan, W. S. Kendall, and J. Mecke, 
\newblock {\it Stochastic Geometry and Its Applications},
\newblock Wiley,
\newblock 2nd ed.,
\newblock 1995.

%
%
%


\bibitem{tse}
D. N. C. Tse and P. Viswanath
\newblock {\it Fundamentals of Wireless Communications},
\newblock Cambridge University Press,
\newblock 2005.

\bibitem{safety}
K. R. Foster, 
\newblock ``A world awash with wireless devices: Radio-frequency exposure issues,"
\newblock {\it IEEE Microwave Mag.},
\newblock vol. 14, 
\newblock no. 2, 
\newblock pp. 73--84,
\newblock 2013.

\bibitem{calka}
P. Calka, 
\newblock ``The distributions of the smallest disks containing the Poisson-Voronoi typical cell and the Crofton cell in the plane,"
\newblock {\it Adv. in Appl. Prob.},
\newblock 702--717,
\newblock 2002.

\bibitem{fca}
H.-S. Jo, Y. J. Sang, P. Xia, and J. G. Andrews,
\newblock ``Heterogeneous cellular
networks with flexible cell association: A comprehensive downlink SINR
analysis,"
\newblock {\it IEEE Trans. Wireless Commun."},
\newblock vol. 11,
\newblock no. 10,
\newblock pp. 3484--3495,
\newblock 2012.

\bibitem{Ram_01}
R. Ramanathan, 
\newblock ``On the performance of ad hoc networks with beamforming antennas." 
\newblock in {\it Proc. ACM international symposium on Mobile ad hoc networking \& computing (MobiHoc)},
\newblock Long Beach,
\newblock United States,
\newblock Oct.
\newblock 2011.

\bibitem{jhpark}
J. Park, S.-L. Kim, and J. Zander, 
\newblock ``Asymptotic Behavior of Ultra-Dense Cellular Networks and Its Economic Impacts,"
\newblock in {\it Proc. IEEE Global Communications Conference (GLOBECOM)},
\newblock Austin,
\newblock United States,
\newblock Dec.
\newblock 2014.


%


\end{thebibliography}
\end{document}